%% file: main.tex
\documentclass[%
 twocolumn,
superscriptaddress,
 amsmath,amssymb,
 aps,
 prd,
]{revtex4-2}
\pdfoutput=1

\usepackage{lipsum}
\usepackage{array}
\usepackage{graphicx}
\usepackage{xcolor}
\usepackage{amsmath}
\usepackage{booktabs}
\usepackage{multirow}
\usepackage{hyperref}
\hypersetup{
	colorlinks=true,
        citecolor=blue
}

\usepackage{natbib}
\setcitestyle{numbers,comma,square}
\usepackage{lineno}

\begin{document}

\title{Detecting Neutrinos from Supernova Bursts in PandaX-4T}
\input{authorlist}

\begin{abstract}
Neutrinos from core-collapse supernovae are essential for the understanding of neutrino physics and stellar evolution. The dual-phase xenon dark matter detectors can provide a way to track explosions of galactic supernovae by detecting neutrinos through coherent elastic neutrino-nucleus scatterings. In this study, a variation of progenitor masses as well as explosion models are assumed to predict the neutrino fluxes and spectra, which result in the number of expected neutrino events ranging from 6.6 to 13.7 at a distance of 10 kpc over a 10-second duration with negligible backgrounds at PandaX-4T. Two specialized triggering alarms for monitoring supernova burst neutrinos are built. The efficiency of detecting supernova explosions at various distances in the Milky Way is estimated. These alarms will be implemented in the real-time supernova monitoring system at PandaX-4T in the near future, providing the astronomical communities with supernova early warnings.
\end{abstract}

\maketitle

\section{Introduction}
\label{introduction}

Massive stars greater than 8 solar masses usually end their lives with core-collapse supernovae (hereafter referred to as SN), but these phenomenons are quite rare and occur only about three times per century in our galaxy~\cite{2009Diffuse}. Each SN explosion ejects a significant amount of material and lasts for about 10 seconds. The total energy released by SN burst is about $10^{53}$ erg, with 99\% carried away by neutrinos~\cite{2020Stellar}. The first recorded SN neutrinos came from SN1987A located in the Large Magellanic Cloud. Approximately 20 neutrino events were detected~\cite{PhysRevLett581494,PhysRevLett581490,ALEXEYEV1988209}, marking the beginning of the neutrino astronomy. These neutrino events provide great insights in the theoretical models of SN explosions, as well as offering constraints to the fundamental properties of neutrinos. However, our present understanding remains inadequate for a thorough comprehension of stellar dynamics and neutrino properties at extreme conditions due to the low statistics of the observed SN neutrino events. Many neutrino experiments and observatories are designed with full readiness to observe the coming SN explosions. 
 
A large number of neutrinos are emitted from the central region of the star at nearly the speed of light during a SN explosion. On the contrary, electromagnetic radiation experiences delayed travel compared to the neutrinos due to the smaller velocity of  outward shock waves~\cite{2016Real}. The delay may vary from several hours to several days, contingent upon SN progenitors. Neutrino observatories located on the Earth can utilize the abrupt rise in the neutrino event rate to provide a prompt alert of the occurrence of a SN explosion in our galaxy to trigger followup optical observations. So far, several experiments have established dedicated SN monitoring systems~\cite{2016Real,2017The,2016Design} and initiated joint observations by the SN early warning system (SNEWS)~\cite{2004SNEWS,AlKharusi2021}.

Freedman pointed out that coherent elastic neutrino-nucleus scattering (CE$\nu$NS) can occur when the incident neutrino energy is on the order of MeV~\cite{PhysRevD91389}. The interaction was initially measured by the COHERENT experiment~\cite{article11}. It implies a large cross section, $\sigma_{\nu}\simeq N^{2}G^{2}_{F}E^{2}_{\nu}/4\pi $, where $N$ represents the neutron number of the target nuclei, $G_F$ is the Fermi constant and $E_{\nu}$ is the neutrino energy~\cite{PhysRevD105043008}. Such a process has been observed in the COHERENT experiment with neutrinos from the spallation neutron sources with an average energy of 50 MeV, providing a sensitive tests on the Standard Model of particle physics. As such, direct detection dark matter experiments using liquid xenon, for example, the operating XENONnT/LUX-ZEPLIN/PandaX-4T experiments~\cite{Aprile2020,LZ2019sgr,PhysRevLett127261802} and the proposed DARWIN~\cite{Aalbers2016} and PandaX-xT experiment~\cite{PandaX:2024oxq}, are good choices to detect the SN neutrinos due to the nucleus's large neutron number of about 80. As the fiducial mass of these experiments increase to above tonne and even ten-tonne level, the CE$\nu$NS process demonstrates a growing competitiveness in the detection of astrophysical neutrinos compared to other neutrino interaction channels. The CE$\nu$NS process is sensitive to all neutrino flavors equally, thereby complementing the studies of experiments that have particular sensitivity to an individual neutrino flavor~\cite{PhysRevD85052007,PhysRevC73035807}. One prominent challenge is that the momentum transfer of the process is very small, resulting in low nuclear recoil (NR) energy at the order of keV, which is difficult to detect. So far, no experiment has observed astrophysical neutrinos via the CE$\nu$NS process.
  
In this paper, the SN model from the Garching group (hereafter referred to as Garching model) is utilized to obtain the neutrino emission spectrum of a SN explosion~\cite{2016NCimR39M,Hdepohl2014NeutrinosFT}. As a comparison, another model proposed by Nakazato et al. is also discussed~\cite{Nakazato2013}. The consideration of neutrino oscillations is unnecessary as we are solely focused on the overall neutrino flux. By taking into account the neutrino flux, cross section, exposure and detection efficiency, the total number of expected neutrinos in PandaX-4T is obtained. A method for accurately predicting the false alert rate is developed. Based on this framework, two specialized triggering alarms, golden and silver, for monitoring SN burst neutrinos are introduced, which can be easily implemented at the software level. Furthermore, the probability of detecting SN explosions in our galaxy out to 50 kpc is estimated. In addition, with the commissioning data acquired from PandaX-4T, an upper limit for the occurrence of SN explosions in the Milky Way is extracted. 
 
The rest of this paper is organized as follows. PandaX-4T experimental setup and data processing are described in Sec.~\ref{sec:experiment}. The detection of SN neutrinos at PandaX-4T is presented in Sec.~\ref{sec:detection}. The trigger algorithm is discussed in Sec.~\ref{sec:algorithm}. The detection probability and the upper limit of SN bursts are shown in Sec.~\ref{sec:limit}.  The conclusions are contained in Sec.~\ref{sec:conclusion}.

\section{PandaX-4T experimental setup and data processing}
\label{sec:experiment}
The PandaX-4T experiment at China Jinping Underground Laboratory (CJPL) has been on operation since November, 2020. The cosmic-ray muons are effectively shielded by a rock overburden of approximately 2,400 meters. The detector is situated within an ultra-pure water shield measuring 13 meters in height and 5 meters in radius. This configuration serves to reduce the impact of the radioactivity that originates from the ambient environment. The detector is a dual-phase xenon time projection chamber (TPC) with about 3.7-tonne liquid xenon  (LXe) in the sensitive volume and 368 Hamamatsu R11410-23 3-inch photo-multipliers (PMTs) which are laid out in a concentric circular (compact hexagonal) for top (bottom) array. And the peripheral background is vetoed by an outer array of 105 Hamamatsu R8520 1-inch PMTs. An upward drift and acceleration electric field for electrons are provided by the three of four transparent stainless steel electrodes. More details can be found at~\cite{Dark}. 

When incident particles pass through liquid xenon, they can generate excited xenon atoms and electron-ion pairs. These particles are detected through the coincidence observation of photon signal from the prompt scintillation (S1) and the delayed electro-luminescence (S2). The photo-electrons (PEs) signals are digitized by the CAEN V1725B digitizer, which operates at a sampling rate of 250 MHz. The digitizer operates in the self-trigger mode, in which any PMT pulse exceeding the threshold of 1/3 PE is recorded. The data are transferred to multiple readout servers and then to a dedicated data-aggregation server in Jingping for further processing and writing to disks. Data are then transferred to servers located in Chengdu for offline physical analysis~\cite{Yang_2022}. The commissioning data from November 28, 2020 to April 16, 2021 with a calendar time of $\sim$86 days are utilized in this work. Detailed operation conditions of the detector and event reconstructions are discussed in~\cite{PhysRevLett127261802}.

\section{Detection of SN neutrinos at PandaX-4T}
\label{sec:detection}
\subsection{Coherent elastic neutrino-nucleus scattering}
\label{subsec:3.1}
\begin{figure}[h] 
\centering
\includegraphics[scale=0.45]{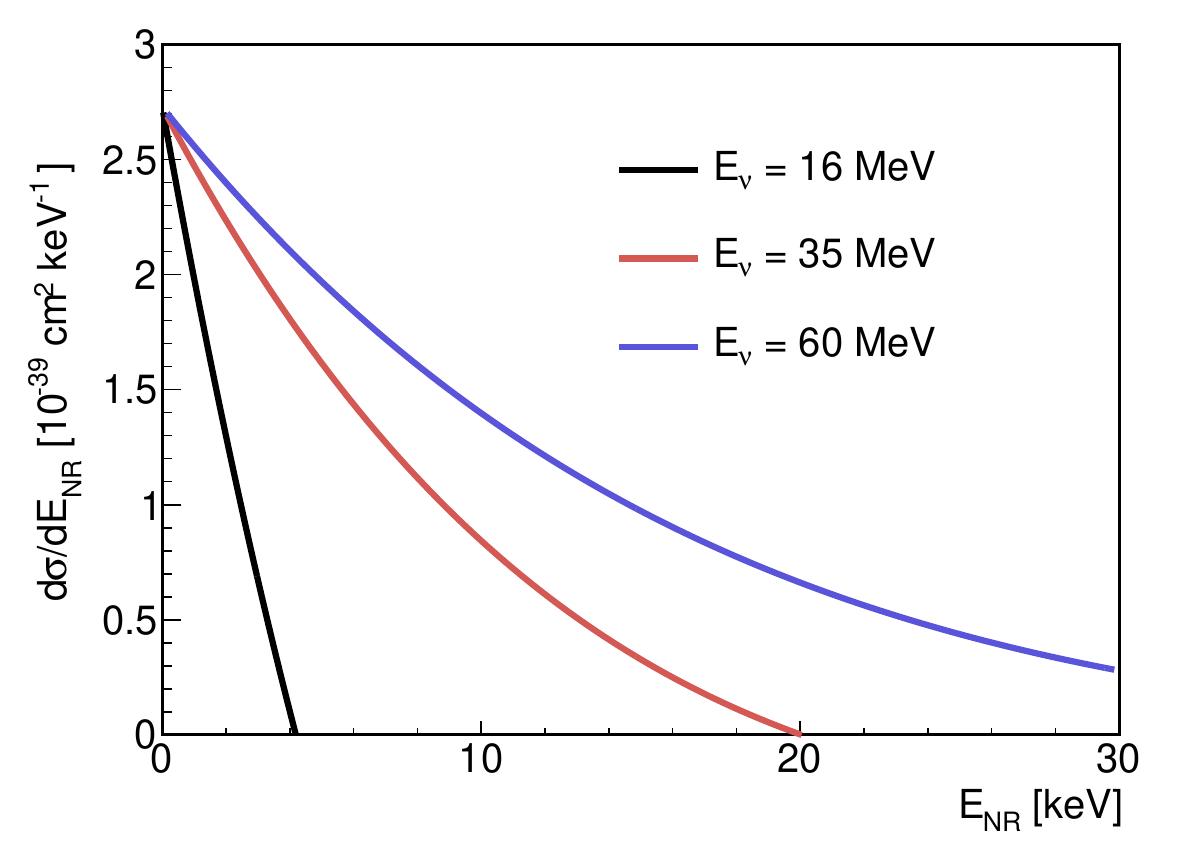}
\caption{\label{fig1} The differential cross section of the CE$\nu$NS process as a function of NR energy. Three different neutrino energies are shown for xenon nuclei. }
\end{figure}

The CE$\nu$NS process can be expressed as:
\begin{equation} \label{eq2}
	\nu_{x}/\bar{\nu}_{x} + A \to \nu_{x}/\bar{\nu}_{x} + A,
\end{equation}
Where $\nu_{x}/\bar{\nu}_{x} ~(x=e, \mu, \tau)$ is any individual flavor of neutrinos/antineutrinos, and $A$ represents the mass number of the target nuclei. The NR energy spans from zero to a maximum value of $E_{max} = 2E^{2}_{\nu}/(2 E_{\nu} + m_{A}) \simeq 2E^{2}_{\nu}/m_{A}$, where $m_{A}$ is the mass of the target nucleus. The differential cross section of the CE$\nu$NS process can be expressed as ~\cite{PhysRevD91389,PhysRevD105043008} 

%\begin{small}
\begin{equation} \label{eq3}
\begin{split}
	\frac{d\sigma}{dE_{N\!R}}(E_{\nu},E_{N\!R}) = &\frac{G^{2}_{F}m_{A}}{4\pi} Q^{2}_{w}(1-\frac{m_{A}E_{N\!R}}{2E^{2}_{\nu}}) \\ 
	& \times F^{2}(q)\Theta(E_{max}-E_{NR}),
\end{split}
\end{equation}
%\end{small}

where ${E_{N\!R}}$ represents the NR energy of the target nucleus, $\Theta(E_{max}-E_{N\!R})$ is the step function with a value of 1 (0) for positive (negative) argument, $Q_{w}$ is the weak charge, 
\begin{equation}
	Q_{w} = N - Z(1-4\text{sin}^{2}\theta_{w}),
	%\nonumber
\end{equation}
where $Z$ is the atomic number of the target nucleus,  $\theta_{w}$ is the weak mixing angle or Weinberg angle with a value of $sin^{2}\theta_{w} \simeq 0.23$~\cite{PhysRevD98030001}, and $F(q)$ is the Helm-type form factor~\cite{PhysRev1041466}, which is defined as
\begin{equation}
	F(q) = \frac{3j_{1}(qr_{0})}{qr_{0}}e^{-\frac{1}{2}(qs)^{2}},
	%\nonumber
\end{equation}
where the momentum-transfer $q=\sqrt{2m_{A}E_{N\!R}}$, $r_{0}= \sqrt{r^2 - 5s^2}$ with the nuclear radius $r=1.2A^{\frac{1}{3}}$ fm and the nuclear skin thickness $s$ of about 0.5 fm, and $j_{1}(qr_{0})$ is the first-order spherical Bessel function~\cite{ENGEL1991114,Kozynets2018SensitivitT}. Figure \ref{fig1} illustrates the differential cross section of Eq.~\eqref{eq3} at different neutrino energies. Note that the CE$\nu$NS exhibits a much larger scattering cross section~\cite{PhysRevD105043008} when compared with inverse beta decays and neutrino-electron elastic scatterings~\cite{article11}, but results in a relatively low NR energy in keV to a several tens of keV. The abundance of xenon isotopes is taken into account in the calculation.

\subsection{Energy spectrum of SN neutrinos}
\begin{figure}[h]
	\centering
	\includegraphics[scale=0.45]{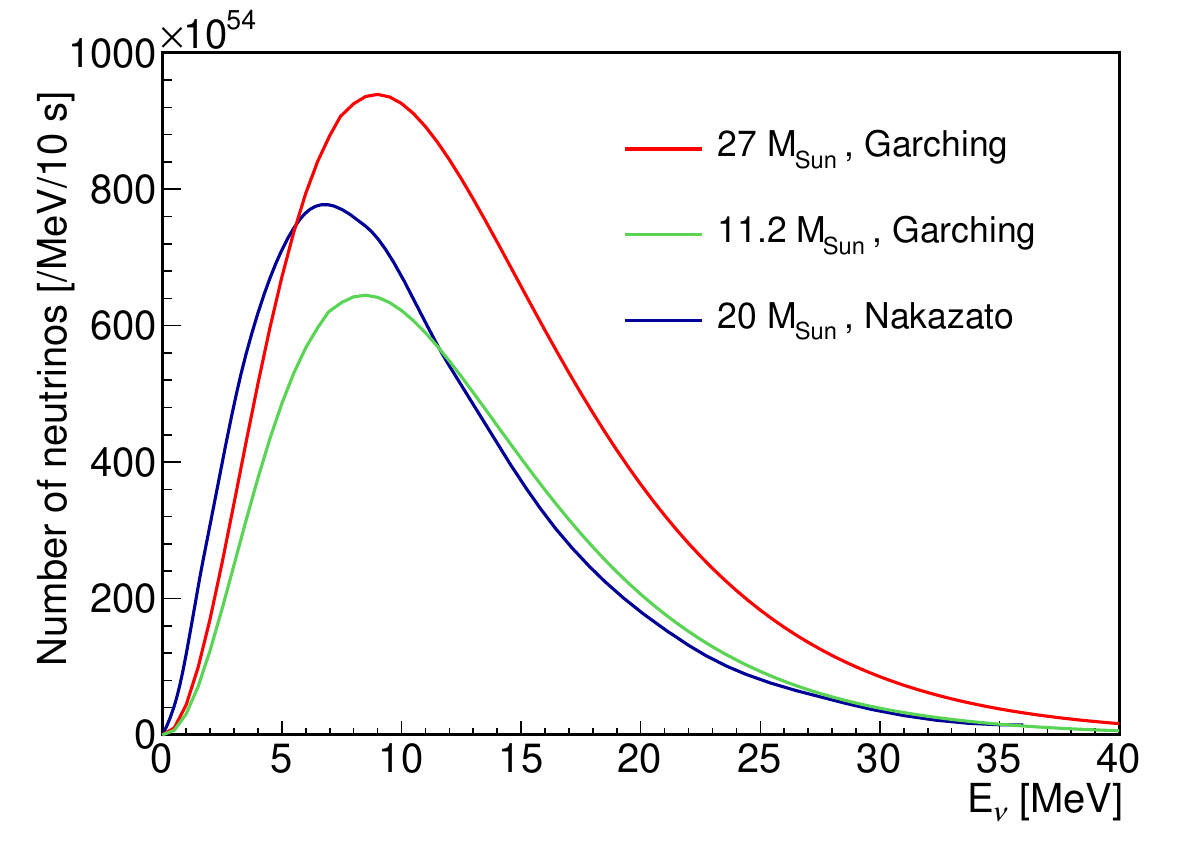}
	\caption{ \label{fig2} The time-integrated energy spectra of all types of SN neutrinos from the core bounce time to the later 10 seconds using Garching and Nakazato models.}
\end{figure}

The SN explosion consists of three stages: the phase of the shock burst, the post-bounce accretion, and the Kelvin-Helmholtz cooling~\cite{2016NCimR39M}. Total emission energy, average energy of neutrinos, and variations in the physical process will lead to different expected fluxes and energy spectra of neutrinos in theoretical models. In this study, we utilize the Garching model as the typical model, where the neutrino energy spectrum can be characterized using the Keil-Raffelt-Janka (KRJ) parametrization~\cite{Keil2003}. The differential flux at a time $t_\mathrm{pb}$, which is defined as the time after the SN core bounce, can be expressed as~\cite{PhysRevD94,PhysRevD97}:
%\begin{large}
\begin{equation} \label{eq4}	
\begin{split}
	\frac{dF(E_{\nu},t_\mathrm{pb})}{dE_{\nu}} =  &\sum_{\nu = 1}^{6}  L_{\nu}(t_\mathrm{pb}) \frac{ ( 1+\gamma (t_\mathrm{pb}) ) ^{1+\gamma (t_\mathrm{pb}) }}{\langle E_{\nu}(t_\mathrm{pb}) \rangle ^{2} \Gamma (1+\gamma (t_\mathrm{pb}) )}  \\
	&(\frac{E_{\nu}}{\langle E_{\nu} (t_\mathrm{pb}) \rangle } ) ^{\gamma(t_\mathrm{pb})} \times \exp[-\frac{(\gamma + 1)E_{\nu}}{ \langle E_{\nu} (t_\mathrm{pb}) \rangle } ]
	 ,
\end{split}
\end{equation}
%\end{large}
where $\nu$ represents one of the six types of neutrinos, $L_{\nu}(t_\mathrm{pb})$ is the neutrino luminosity, $\langle E_{\nu}(t_\mathrm{pb}) \rangle $ is the mean energy of neutrinos at time $t_\mathrm{pb}$ and $\Gamma(1+\gamma (t_\mathrm{pb}))$ is the Gamma function. The spectral index $\gamma(t_\mathrm{pb})$ can be obtained as~\cite{PhysRevD94}
\begin{equation}
	\frac{\langle E_{\nu}(t_\mathrm{pb}) ^{2} \rangle}{\langle E_{\nu}(t_\mathrm{pb}) \rangle ^{2}} = \frac{2 + \gamma (t_\mathrm{pb}) }{1+ \gamma (t_\mathrm{pb})}. 
	%\nonumber
\end{equation}
in which $\langle E_{\nu}(t_\mathrm{pb}) ^{2} \rangle$ represents the mean of $E_{\nu}(t_\mathrm{pb}) ^{2}$. Here, we use models with two benchmark progenitor masses $M_{p}=11.2~M_{\odot}$ and $M_{p}=27~M_{\odot}$ ($M_{\odot}$ is the solar mass), employing the LS220 nuclear equation of state (EoS)~\cite{LATTIMER1991331}, to predict the neutrino fluxes and the energy spectra. As a contrast, the Nakazato model with $M_{p}=20~ M_{\odot}$, metallicity Z=0.02, and shock revival time $t_{rev}=200 $ ms is also used~\cite{Nakazato2013}. Figure~\ref{fig2} displays the time-integrated energy spectra of all types of neutrinos from the core bounce time to the later 10 seconds. For the purpose of clear visual comparison between the models, neutrino energies are shown only up to 40 MeV. Considering that the neutrino flux from a SN explosion is predominantly concentrated within the first few seconds after the core bounce, we have approximated the integration time for several models to be 10 seconds. The fraction of neutrino flux beyond 10 seconds is negligible. 

\subsection{Observable in liquid xenon detectors}
\label{sec:3.3}
When neutrinos interact with xenon atoms through CE$\nu$NS process, the differential event rate can be described as
\begin{equation} \label{eq5}
\begin{split}
	\frac{dN_{0}}{dE_{N\!R}}(E_{N\!R})=&\frac{m_{det}N_{A}}{M_{A}(4\pi D^{2})} \int_{E^{min}_{\nu}}^{\infty}	 \frac{d\sigma}{dE_{N\!R}}(E_{\nu},E_{N\!R}) \\
    &\times f(E_{\nu})dE_{\nu},
 \end{split}
\end{equation}
where $m_{det}=2.67$ tonnes is the effective mass of liquid xenon in this work, and $N_{A}$ denotes the Avogadro's constant, $M_{A}$ represents the molar mass of xenon atoms, and the expression $m_{det} N_{A}/ M_{A}$ is the total number of xenon atoms in the detector. $D$ is the distance from Earth to the SN. In order to produce a recoil energy above $E_{N\!R}$, it is mandatory to have a minimum neutrino energy, which can be expressed as $E^{min}_{\nu} = \sqrt{m_{A} E_{N\!R}/2}$ and is the lower limit of the integral in Eq.~\eqref{eq5}. $f(E_{\nu})$ is the energy spectrum for the sum of all neutrino types, as depicted in Fig. \ref{fig2}. Practically, it is enough to integrate over $E_{\nu} \in (E^{min}_{\nu},100$ MeV). Considering the detection efficiency $\epsilon(E_{N\!R})$, the effective differential event rate is~\cite{ABE201751}
\begin{equation} \label{eq6}
	\frac{dN}{dE_{N\!R}}(E_{N\!R})= \epsilon (E_{N\!R}) \times \frac{dN_{0}}{dE_{N\!R}}(E_{N\!R}),
\end{equation}

In order to obtain the detection efficiency, the Monte Carlo (MC) simulation is performed~\cite{WF_2023}. Firstly, the light yield ($L_{y}$) and charge yield ($Q_{y}$) of NR events in liquid xenon are studied using the Noble Element Simulation Technique (NEST) v2.3.6 parameterization~\cite{2018Noble,instruments5010013}, with parameters tuned by the calibrations as described  in~\cite{PhysRevLett127261802,PhysRevLett130021802}. In the simulation, we have taken into account various detector effects, including the quantum effects of PMTs and the non-uniformity of the detector, etc. $2 \times 10^{5}$ events are simulated for each mono-energetic point ranging from 0.1 to 30 keV, with a step size of 0.1 keV. Subsequently, $L_{y}$ and $Q_{y}$ are used as the inputs for the waveform simulation (WS)~\cite{WF_2023}, which is capable of generating complete waveforms. To mimic the detector response of the simulated events, the effects of afterpulsing from PMTs and delayed electrons are considered in the WS process.  

\begin{figure}[h]
	\centering
	\includegraphics[scale=0.45]{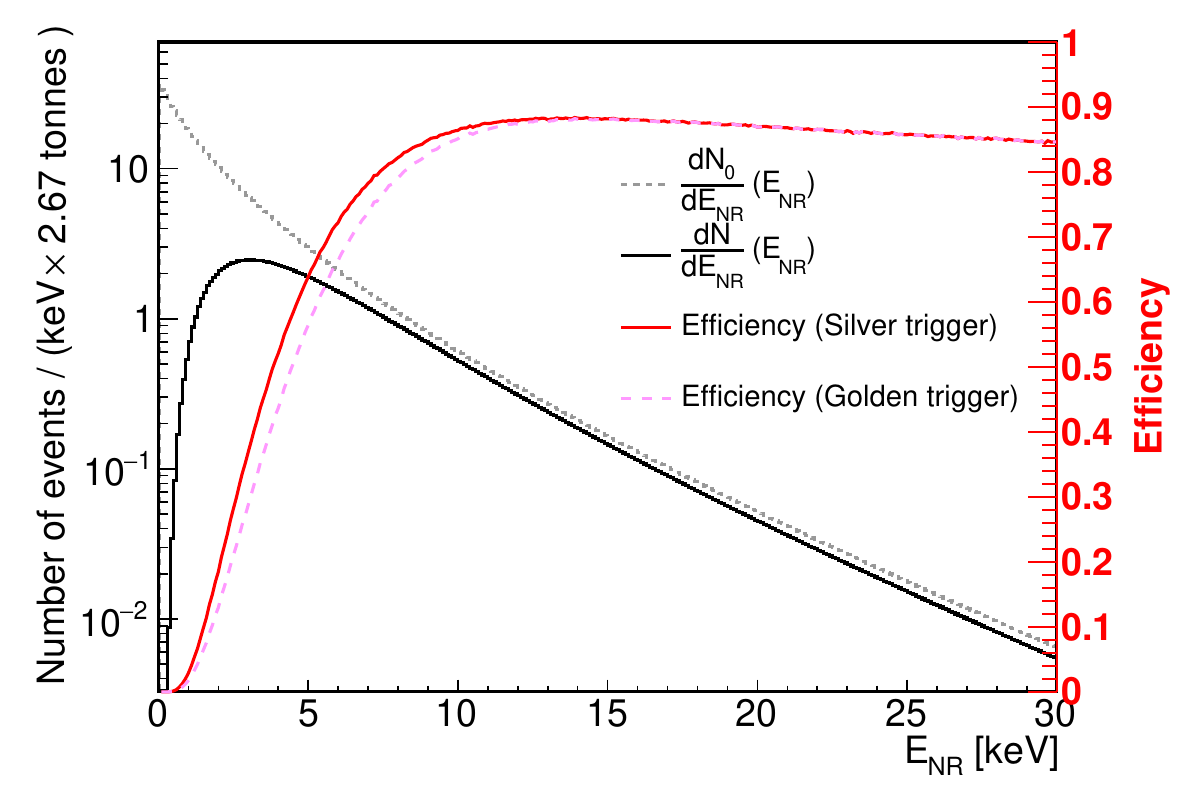}
	\caption{\label{fig3} The detection efficiencies of silver trigger (red solid) and golden trigger(magenta dashed) as a function of NR energy, the expected neutrino spectrum before (gray) and after (black) the efficiency (red solid) correction. The Garching model with a progenitor mass of $27~M_{\odot}$ and the LS220 EoS is used. The distance from SN to Earth is assumed to be 10 kpc. The neutrino energy spectrum is integrated from the onset time of the core bounce to the subsequent 10 seconds.}
\end{figure}

The simulated events are processed with similar procedures as in the experimental data. The total efficiency primarily consists of three parts: the signal reconstruction, the data quality selection, and the region-of-interest (ROI). The preceding two items were discussed in detail in the prior analysis~\cite{PhysRevLett130021802}. To improve the detection efficiency, we loosen several data quality selection cuts and adjust the ROI. Two ROIs are used, one corresponds to the golden alert which is stricter to trigger SN alarms than the silver one. The former (latter) is defined to have S1 ranging from 2.1 to 100 PEs (1.65 to 100 PEs) and S2 ranging from 80 to 3500 PEs (same). This is detailed in Sec.~\ref{sec:algorithm}. The magenta dashed (red solid) line depicted in Fig. \ref{fig3} illustrates the detection efficiency as a function of NR energy in the case of the golden (silver) alarm. A decrease in efficiency occurs at energies above 15 keV, which is attributed to the diffusion cut on the drift electrons in the selection process. The total detection efficiency is $20\%$ and $23\%$ for the golden (silver) alarm in the case of Garching model with $M_{p}=27~M_{\odot}$, and other two SN models used in this paper have the similar efficiencies. Note that in order to maximize the detection efficiency, the NR/ER cut based on the ratio between the number of the ionized electrons and the photons is not used in this paper. And this can be revived in the PandaX-nT experiment due to the much larger target mass.    
%Our primary focus is directed toward the lower energy range. It is important to note that this encompasses the efficiency of NEST's reconstruction process. 

The NR spectrum of CE$\nu$NS before (gray) and after (black) the efficiency (red solid) correction in liquid xenon is illustrated in Fig.~\ref{fig3}. The total expected number of observable neutrinos $N_{obs}$, which is mainly from the high-energy neutrinos (approximately 15 to 30 MeV) as illustrated in Fig.~\ref{fig2}, can be expressed as:
\begin{equation} \label{eq7}
  	N_{obs}= \int \frac{dN}{dE_{N\!R}}(E_{N\!R}) dE_{N\!R},
\end{equation}
here, we perform the integration up to 30 keV since the flux above has a negligible impact on $N_{obs}$. In order to increase $N_{obs}$, a high detection efficiency is quite desired. The numbers of SN neutrinos using the Garching and the Nakazato models at two different distances of 10 kpc and 168 pc are listed in Table \ref{tab:l1}, where 10 kpc (168 pc) is approximately the distance from the center of the Milky Way (Betelgeuse) to Earth. The background event rate, as listed in Table~\ref{tab:l2}, is negligible compared with the event rate of the SN signals as listed in Table~\ref{tab:l1}. Betelgeuse is considered as a potential candidate for a SN explosion~\cite{Joyce2020}, which would provide an effective handle to  distinguish between different models by observing the number and spectrum of NR events due to the short distance. In the case of Betelgeuse explosion, the design of the data acquisition needs to address the challenge of high data rate during the explosion. 

\renewcommand\arraystretch{1.5}
\begin{table}[htbp]
	\setlength{\abovecaptionskip}{0cm}
	\setlength{\belowcaptionskip}{0.2cm}
	\centering
	\caption{\label{tab:l1} The number of expected SN neutrinos from 10 kpc and 168 pc in PandaX-4T in the case of golden and silver alarms. Two Garching models are used with $M_{p}=11.2~M_{\odot}$ and $M_{p}=27~M_{\odot}$, employing the LS220 EoS. Nakazato model with $M_{p}=20~M_{\odot}$, Z (metallicity)$=0.02$, and $t_{rev}$ (the shock revival time) $=200$ ms is used for comparison. The neutrino energy spectrum is integrated from the onset time of the core bounce to the subsequent 10 seconds.}
	\setlength{\tabcolsep}{0.4mm}{
	\begin{tabular}{ccccc p{0.15cm}<{\centering}p{0.15cm}<{\centering}p{0.15cm}<{\centering}p{0.15cm}<{\centering}p{0.15cm}<{\centering} }    
	    \toprule[0.4mm] & \\ [-4.3ex]	   
	    \hline
		\multirow{2}*{SN model} & \multicolumn{2}{c}{Golden alarm} & \multicolumn{2}{c}{Silver alarm} & \\	
		\cmidrule(lr){2-3}\cmidrule(lr){4-5}
		& D=10 kpc & 168 pc & 10 kpc & 168 pc \\
		\midrule	
		20 $M_{\odot}$ Nakazato& 7.2 & 2.6$\times$ $10^{4}$  & 8.3 & 2.9$\times$ $10^{4}$  \\
		11.2 $M_{\odot}$ Garching & 6.6 & 2.3$\times$ $10^{4}$ & 7.7 & 2.7$\times$ $10^{4}$  \\
		27 $M_{\odot}$ Garching & 13.7 & 4.9$\times$ $10^{4}$ & 15.9 & 5.7$\times$ $10^{4}$ \\	
		\hline &\\ [-4.ex]
		\bottomrule[0.4mm]
	\end{tabular} }
\end{table}

\section{SN neutrino trigger algorithm}
\label{sec:algorithm}
\subsection{Trigger algorithm}
\label{sec:4.1}
The explosion of a SN would result in a sudden increase in the event rate within a short period of several seconds in the detector, providing a unique opportunity to observe this phenomenon in the Milky Way. The software-based SN trigger can provide a method for monitoring SN explosions as soon as data files are written into the disk clusters. The trigger algorithm consists of three parts 1) event builder, 2) selection of signal candidates, and 3) SN neutrino trigger. The event builder process involves clustering PMT hits into signal pulses and classifying these pulses into S1s and S2s, where the classified S1s and S2s are paired to build incident events.  The size of each file is significantly decreased after the event builder process, from 1 gigabyte (GB) to approximately 100 megabytes (MB). During the commissioning phase, the background event rate is stable. To select good event candidates, those basic cuts derived from the analysis of the solar $^8$B study~\cite{PhysRevLett130021802} are used to suppress background events. The events that survive the cuts are used for the investigation of the false alert rate, which will be discussed in Sec.~\ref{sec:false}. For the SN neutrino trigger, when the trigger algorithm identifies the first candidate, it will serve as the starting point of the search time. Subsequently, the total number of events is counted within the following ten-second time window. If the counted number exceeds the specific threshold $N_{thr}$, a prompt alert will be issued. Just to be clear, if another candidate appears outside of the time window, it will serve as a new starting point of the search time, and the same process continues. When an alert is issued, the information including the start time of the alert and the total number of candidates in the time window are sent simultaneously to the experts of the PandaX-4T SN group by e-mail. The relevant data files will be stored in the designated directory for further examination by the experts. The whole process takes several minutes for each individual file on average. 
  
\subsection{False alert}
\label{sec:false}
\begin{figure}[h]
	\centering
	\includegraphics[scale=0.75]{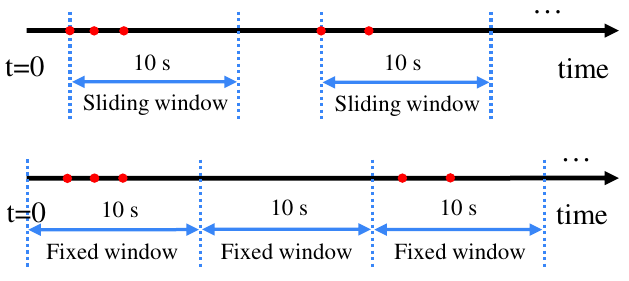}
	\caption{\label{fig4} Schematic of sliding and fixed time window. The red dots represent SN neutrino candidates. }
\end{figure}

\begin{figure}[h]
	\centering
	\includegraphics[scale=0.45]{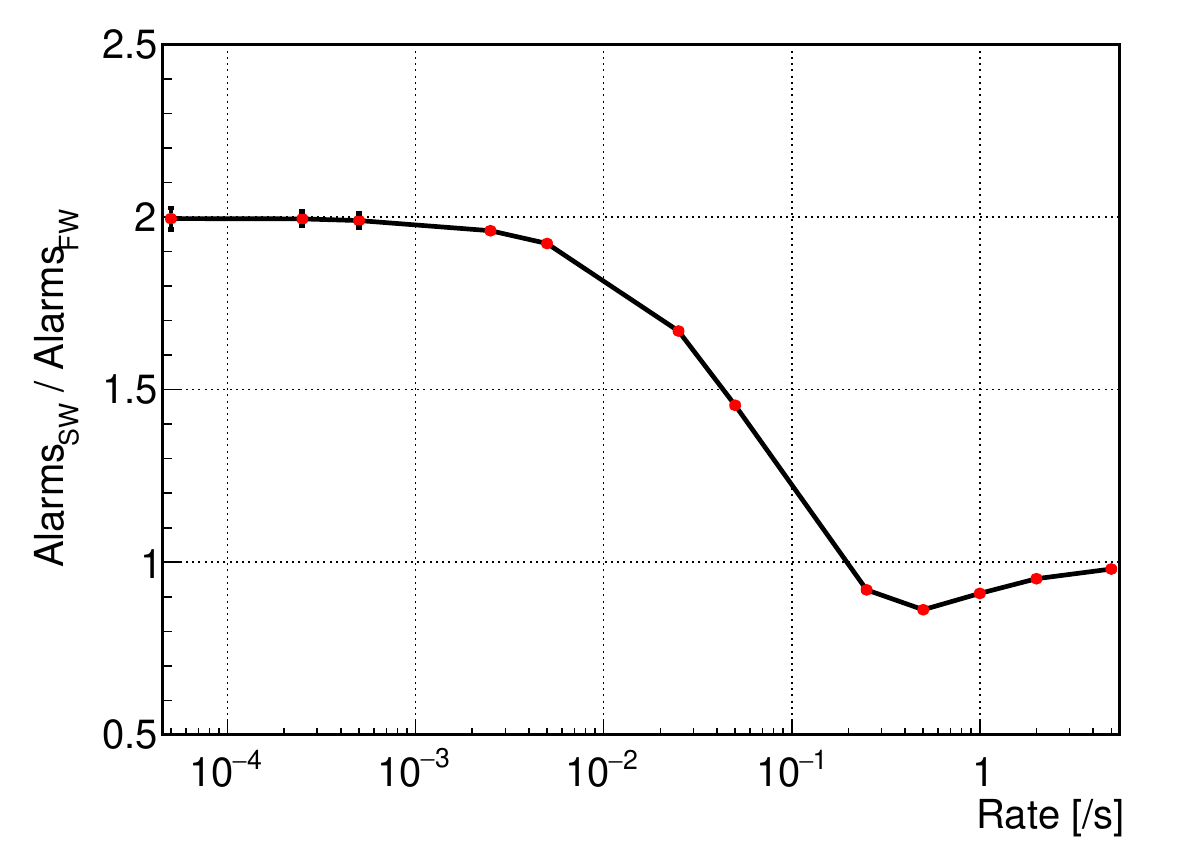}
	\caption{\label{fig5} The ratio of the false alerts using sliding and fixed time window. The parameters used are $T_{SN} =10$ s, $N_{thr} =2$ and the uncertainty arises from the estimation of the mean values. The curve is connected through the data points via the linear interpolation, and the data points are calculated from the toy MC simulations. }
\end{figure}

It is assumed that the number of events observed by the detector follows a Poisson distribution. The probability for the number of events in a time window $T_{SN}$ no less than the threshold value $N_{thr}$ follows
\begin{equation} \label{eq8}
	p(N_{thr};T_{SN};r_{bg}) = 1 - \sum_{n=0}^{N_{thr} -1} \frac{1}{n!} e^{-r_{bg}T_{SN}} (r_{bg}T_{SN})^{n},
\end{equation}
where $r_{bg}$ is the mean background event rate. The false alert rate per week using a fixed time window $T_{SN}$ can be written as
\begin{equation} \label{eq9}
	R_{false} = \frac{3600  \cdot 24 \cdot 7 }{T_{SN}} p(N_{thr};T_{SN};r_{bg}),
\end{equation}
The backgrounds are mainly composed of material radioactivity, environmental radioactivity, contamination from tritium calibration, and additional radioactive sources in xenon such as $^{222}$Rn~\cite{Dark}. In order to improve the SN detection probability while simultaneously reducing the false alert rate, we investigate various combinations of $T_{SN}$ and $N_{thr}$ at a certain event rate $r_{bg}$. Eventually, a suitable combination of the values is chosen, with $T_{SN}=10$ s and $N_{thr} =2$. Unless stated otherwise, we will always use this default setting. The false alert rate is predicted with a two-step procedure. In the first step, one can calculate the number of false alert rate for a given event rate $r_{bg}$ and time window using Eq. \eqref{eq9}. Note this number is for the case of the fixed time window (i.e., the subsequent time window is seamlessly connected to the previous time window, so there is no time gap between any two time windows), and in our algorithm we use the sliding time window (i.e., the time of the candidate is used as the start of the time window, and there could be a time gap between two neighbour time windows). A schematic of sliding and fixed time window is shown in Fig.~\ref{fig4}. In the second step, the ratio of the number of false alert rate with the sliding and fixed time window is calculated based on a toy MC simulation. The former (latter) is referred to as SW (FW) method hereafter. In the toy MC simulation, events are randomly sampled assuming a specific event rate, followed by the counting of the total number of false alerts with the SW and FW methods. To reduce the uncertainty of the mean ratio value, hundreds of simulations are carried out for each event rate. Figure \ref{fig5} shows the ratio as a function of the event rate. 

It is necessary to emphasize three features in Fig.~\ref{fig5}. Firstly, when the rate is less than $10^{-3}$ /s, the ratio gradually converges to 2. A mathematical explanation can be provided based on the the settings for this study. At the low rate, each alert comprises only two candidates with the parameters of $T_{SN}=10$ s and $N_{thr} =2$, due to the fact that the probability of greater than two candidates appearing within 10 seconds is significantly lower than the case of exact two candidates. Supposing two candidates occur within a 10-second window in a total time length which can be divided into N intervals, with each interval of 5 seconds and N being sufficiently large, there are 3N/2 and  N/2 scenarios in the case of two candidates occurring in different intervals for SW and FW methods, respectively, while the number of the scenarios in the case of two candidates occurring in same intervals are both N/2. As a result, the ratio is (3N/2+N/2)/(N/2+N/2)=2. The ratio of 2 can also be explained as ($r_{bg}\times10$ s)/($r_{bg}\times5$ s), since the first candidate appears at the start (middle on average) of a 10-second window for SW (FW) method. Note that the number of false alerts expected from Eq.~\eqref{eq9} is in good agreement with the number counted from the toy MC simulation using FW method. In contrast to the SW method, the FW method divides the time length into uniform intervals according to the time window $T_{SN}$ and counts the false alerts within each interval. Secondly, the ratio reduces to a value below 1 due to the fact that with higher rate, the possibility of more than two CE$\nu$NS events piling up in a single sliding window grows, leading to a less number of triggers. Thirdly, the ratio converges to 1 at high event rate great than 5 /s, since two methods show negligible difference in these cases.  

\subsection{Validation of false alert rate}
To validate the false alert rate described above, experimental data including the deuteron-deuteron (DD),  AmBe neutron calibration and the commissioning physical data are used to test the trigger algorithm. Eq. \eqref{eq9} enables the inverse calculation of the corresponding background event rate for a given false alert rate. As stated previously in Sec.~\ref{sec:3.3}, two trigger modes (golden and silver) for the commissioning physical data have been chosen, which correspond to the false alert rate (background event rate) of one per month ($\sim2.2\times10^{-4}$ /s) and one per week ($\sim3.6\times10^{-4}$ /s), respectively. The ROI plays a significant role in regulating the false alert rate, which is in turn used as a reference to quickly adjust the ROI in the commissioning physical data. The numbers of the expected false alerts are consistent with the observed, as listed in Table \ref{tab:l2}. For the purpose of simplicity, only the results with silver trigger mode are shown. The scenario where the two candidates fall into two separate files has been considered. 
\renewcommand\arraystretch{1.5}
\begin{table}[htbp]
	\setlength{\abovecaptionskip}{0cm}
	\setlength{\belowcaptionskip}{0.2cm}
	\setlength{\tabcolsep}{3pt}
	\centering
	\caption{\label{tab:l2}Comparison of the observed and the expected false alerts with the silver trigger mode using the DD, AmBe and physical data as listed in fourth and fifth column. The second and third column list the event rate and the calendar time of the dataset. }

	\begin{tabular}{ccccc p{2.cm}<{\centering}p{2.cm}<{\centering}p{2.cm}<{\centering}p{2.cm}<{\centering}p{2.cm}<{\centering}}
		\toprule[0.4mm] &\\ [-4.3ex]	   
		\hline
		Data type & Rate [/s] & Calendar time & Observed & Expected \\
		\hline
		DD& 3.61$\times$$10^{-3}$ & 3.58 days & 40 & 38 \\
		AmBe& 3.12$\times$$10^{-3}$ & 5.7 days & 49 & 46 \\
		Physical & 3.6$\times$$10^{-4}$ & 86.1 days & 8 & 9.8 \\	
		\hline &\\ [-4.ex]
		\bottomrule[0.4mm]
	\end{tabular}
\end{table}

\section{Detection probability and upper limit of SN bursts}
\label{sec:limit}
\begin{figure}[h]
	\centering
	\includegraphics[scale=0.45]{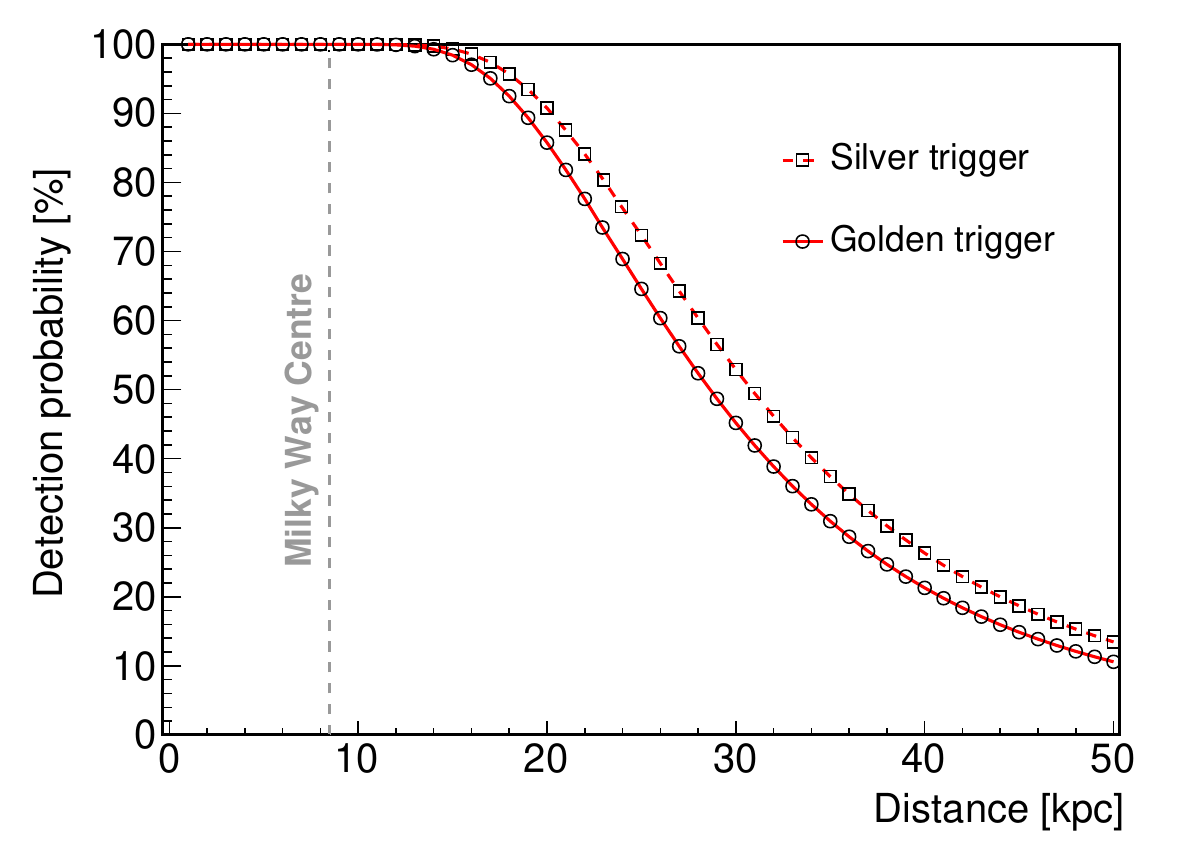}
	\caption{\label{fig6} The detection probability of the SN explosion as a function of distance from the SN to the Earth. The Garching model is used with $M_{p}=27~M_{\odot}$, employing the LS220 EoS. }
\end{figure}

The detection probability of the SN explosions is same as~Eq.\eqref{eq8}, with the change of  $r_{bg}$ to $r_{SN}$, where $r_{SN}$ is the event rate of SN neutrinos. Figure \ref{fig6} illustrates the probability of detecting SN explosions at different distances using the golden and silver trigger modes with the Garching model assuming the progenitor mass of $M_{p}=27~M_{\odot}$. The PandaX-4T detector would have a 100$\%$ probability of detecting SN explosions with the golden and silver trigger modes assuming the SN is located 10 kpc from the earth.

It is predicted that 3.4 false alerts would happen in the physical data with the golden trigger mode, and one is observed. The $p$-value for this downward fluctuation is 14.7\%. An upper limit at 90\% confidence level (C.L.) is extracted to be 1.6 using the Feldman-Cousin method~\cite{Feldman:1997qc}, which is equivalent to 678.2 per century on the rate of core-collapse supernovae in our galaxy out to 10 kpc. 

\section{Conclusions}
\label{sec:conclusion}
The NR spectra of CE$\nu$NS process from SN neutrinos are investigated in PandaX-4T, leading to the expected number of neutrino events ranging from 6.6 to 13.7 by varying the mass of the progeny stars and the explosion models at a distance of 10 kpc with a time duration of 10 seconds in the case of the golden trigger mode. The uncertainties in the parameters of the SN theoretical models will be greatly reduced once future SN explosions are observed by the detectors on the Earth. Neutrino detectors can provide an early warning for the multi-messenger astronomy, which can better understand the SN dynamics from different aspects. The algorithm to predict the false alert rate is well established and validated by the experimental data. It takes several minutes to search for SN bursts for each individual file and will take a bit longer if the case that candidates falling into two separate files is taken into account. The detection probability for SN explosions within the Milky Way is given, which suggests that the PandaX-4T detector is capable of monitoring SN explosions in the Milky Way. The upper limit of SN bursts at 90\% C.L. is determined to be 678.2 per century in our galaxy out to 10 kpc.  In the future, the trigger algorithm will be integrated into the online SN monitor system in PandaX-4T, providing a real-time monitoring of SN bursts in our galaxy.  

%%cite{sweere2022deep} 

\section*{Acknowledgements}
This project is supported in part by the grants from National Science Foundation of China (Nos. 12090060, 12090063, 12105052, 12005131, 11905128, 11925502), and by the Office of Science and Technology, Shanghai Municipal Government (grant No. 22JC1410100). We thank the support from the Double First Class Plan of Shanghai Jiao Tong University. We also thank the sponsorship from the Chinese Academy of Sciences Center %for Excellence in Particle Physics (CCEPP), Hongwen Foundation in Hong Kong, Tencent, and New Cornerstone Science Foundation in China. 
for Excellence in Particle Physics (CCEPP), Hongwen Foundation of Hong Kong, Tencent, New Cornerstone Science Foundation, and Discipline Construction Fund of Shandong University in China. Finally, we thank the CJPL administration and the Yalong River Hydropower Development Company Ltd. for indispensable logistical support and other help. 

%% The Appendices part is started with the command \appendix;
%% appendix sections are then done as normal sections
%% \label{}

%% If you have bibdatabase file and want bibtex to generate the
%% bibitems, please use
%%
%%\bibliographystyle{elsarticle-harv} 

 %%\bibliographystyle{plainnat} 
\bibliographystyle{unsrt}
\bibliography{main.bib}

%\bibliographystyle{apsrev4-1}
%\bibliography{ref.bib}
\end{document}

%% file: authorlist.tex
% !TEX root = ../main.
\def\TDLee{New Cornerstone Science Laboratory, Tsung-Dao Lee Institute, Shanghai Jiao Tong University, Shanghai, 200240, China}
\def\shKeyLab{School of Physics and Astronomy, Shanghai Jiao Tong University, Key Laboratory for Particle Astrophysics and Cosmology (MoE), Shanghai Key Laboratory for Particle Physics and Cosmology, Shanghai 200240, China}
\def\BUAA{School of Physics, Beihang University, Beijing 102206, China}
\def\BUAALab{Beijing Key Laboratory of Advanced Nuclear Materials and Physics, Beihang University, Beijing, 102206, China}
\def\zzu{School of Physics and Microelectronics, Zhengzhou University, Zhengzhou, Henan 450001, China}
\def\USTClab{State Key Laboratory of Particle Detection and Electronics, University of Science and Technology of China, Hefei 230026, China}
\def\USTCdep{Department of Modern Physics, University of Science and Technology of China, Hefei 230026, China}
\def\BUAALab{International Research Center for Nuclei and Particles in the Cosmos \& Beijing Key Laboratory of Advanced Nuclear Materials and Physics, Beihang University, Beijing 100191, China}
\def\pku{School of Physics, Peking University, Beijing 100871, China}
\def\YaLongSD{Yalong River Hydropower Development Company, Ltd., 288 Shuanglin Road, Chengdu 610051, China}
\def\IAP{Shanghai Institute of Applied Physics, Chinese Academy of Sciences, 201800 Shanghai, China}
\def\CHEPpku{Center for High Energy Physics, Peking University, Beijing 100871, China}
\def\SDUdep{Research Center for Particle Science and Technology, Institute of Frontier and Interdisciplinary Science, Shandong University, Qingdao 266237, Shandong, China}
\def\SDUlab{Key Laboratory of Particle Physics and Particle Irradiation of Ministry of Education, Shandong University, Qingdao 266237, Shandong, China}
\def\UMD{Department of Physics, University of Maryland, College Park, Maryland 20742, USA}
\def\MESJTU{School of Mechanical Engineering, Shanghai Jiao Tong University, Shanghai 200240, China}
\def\SYU{School of Physics, Sun Yat-Sen University, Guangzhou 510275, China}
\def\SYUSFI{Sino-French Institute of Nuclear Engineering and Technology, Sun Yat-Sen University, Zhuhai, 519082, China}
\def\NKU{School of Physics, Nankai University, Tianjin 300071, China}
\def\YTU{Department of Physics,Yantai University, Yantai 264005, China}
\def\FDU{Key Laboratory of Nuclear Physics and Ion-beam Application (MOE), Institute of Modern Physics, Fudan University, Shanghai 200433, China}
\def\USST{School of Medical Instrument and Food Engineering, University of Shanghai for Science and Technology, Shanghai 200093, China}
\def\SJTUSC{Shanghai Jiao Tong University Sichuan Research Institute, Chengdu 610213, China}
\def\SPEIT{SJTU Paris Elite Institute of Technology, Shanghai Jiao Tong University, Shanghai, 200240, China}
\def\NNU{School of Physics and Technology, Nanjing Normal University, Nanjing 210023, China}
\def\SYUzhuhai{School of Physics and Astronomy, Sun Yat-Sen University, Zhuhai, 519082, China}

\affiliation{\TDLee}
\author{Binyu Pang}\affiliation{\SDUdep}\affiliation{\SDUlab}
\author{Abdusalam Abdukerim}\affiliation{\shKeyLab}
\author{Zihao Bo}\affiliation{\shKeyLab}
\author{Wei Chen}\affiliation{\shKeyLab}
\author{Xun Chen}\affiliation{\SJTUSC}
%\author{Yunhua Chen}\affiliation{\YaLongSD}
\author{Chen Cheng}\affiliation{\SYU}
\author{Zhaokan Cheng}\affiliation{\SYUSFI}
\author{Xiangyi Cui}\affiliation{\TDLee}
\author{Yingjie Fan}\affiliation{\YTU}
\author{Deqing Fang}\affiliation{\FDU}
\author{Changbo Fu}\affiliation{\FDU}
\author{Mengting Fu}\affiliation{\pku}
\author{Lisheng Geng}\affiliation{\BUAA}\affiliation{\BUAALab}\affiliation{\zzu}
\author{Karl Giboni}\affiliation{\shKeyLab}
\author{Linhui Gu}\affiliation{\shKeyLab}
\author{Xuyuan Guo}\affiliation{\YaLongSD}
\author{Chencheng Han}\affiliation{\TDLee} 
\author{Ke Han}\affiliation{\shKeyLab}
\author{Changda He}\affiliation{\shKeyLab}
\author{Jinrong He}\affiliation{\YaLongSD}
\author{Di Huang}\affiliation{\shKeyLab}
\author{Yanlin Huang}\affiliation{\USST}
\author{Junting Huang}\affiliation{\shKeyLab}
\author{Zhou Huang}\affiliation{\shKeyLab}
\author{Ruquan Hou}\affiliation{\SJTUSC}
\author{Yu Hou}\affiliation{\MESJTU}
\author{Xiangdong Ji}\affiliation{\UMD}
\author{Yonglin Ju}\affiliation{\MESJTU}
\author{Chenxiang Li}\affiliation{\shKeyLab}
\author{Jiafu Li}\affiliation{\SYU}
\author{Mingchuan Li}\affiliation{\YaLongSD}
%\author{Shu Li}\affiliation{\MESJTU}
\author{Shuaijie Li}\affiliation{\TDLee}
\author{Tao Li}\affiliation{\SYUSFI}
\author{Qing Lin}\affiliation{\USTClab}\affiliation{\USTCdep}
\author{Jianglai Liu}\email[Spokesperson: ]{jianglai.liu@sjtu.edu.cn}\affiliation{\TDLee}\affiliation{\shKeyLab}\affiliation{\SJTUSC}
\author{Congcong Lu}\affiliation{\MESJTU}
\author{Xiaoying Lu}\affiliation{\SDUdep}\affiliation{\SDUlab}
\author{Lingyin Luo}\affiliation{\pku}
\author{Yunyang Luo}\affiliation{\USTCdep}
\author{Wenbo Ma}\affiliation{\shKeyLab}
\author{Yugang Ma}\affiliation{\FDU}
\author{Yajun Mao}\affiliation{\pku}
\author{Yue Meng}\affiliation{\shKeyLab}\affiliation{\SJTUSC}
\author{Xuyang Ning}\affiliation{\shKeyLab}
\author{Ningchun Qi}\affiliation{\YaLongSD}
\author{Zhicheng Qian}\affiliation{\shKeyLab}
\author{Xiangxiang Ren}\affiliation{\SDUdep}\affiliation{\SDUlab}
\author{Nasir Shaheed}\affiliation{\SDUdep}\affiliation{\SDUlab}
%\author{Changsong Shang}\affiliation{\YaLongSD}
\author{Xiaofeng Shang}\affiliation{\shKeyLab}
\author{Xiyuan Shao}\affiliation{\NKU}
\author{Guofang Shen}\affiliation{\BUAA}
\author{Lin Si}\affiliation{\shKeyLab}
\author{Wenliang Sun}\affiliation{\YaLongSD}
\author{Andi Tan}\affiliation{\UMD}
\author{Yi Tao}\affiliation{\shKeyLab}\affiliation{\SJTUSC}
\author{Anqing Wang}\affiliation{\SDUdep}\affiliation{\SDUlab}
\author{Meng Wang}\affiliation{\SDUdep}\affiliation{\SDUlab}
\author{Qiuhong Wang}\affiliation{\FDU}
\author{Shaobo Wang}\affiliation{\shKeyLab}\affiliation{\SPEIT}
\author{Siguang Wang}\affiliation{\pku}
\author{Wei Wang}\affiliation{\SYUSFI}\affiliation{\SYU}
\author{Xiuli Wang}\affiliation{\MESJTU}
\author{Zhou Wang}\affiliation{\shKeyLab}\affiliation{\SJTUSC}\affiliation{\TDLee}
\author{Yuehuan Wei}\affiliation{\SYUSFI}
\author{Mengmeng Wu}\affiliation{\SYU}
\author{Weihao Wu}\affiliation{\shKeyLab}
\author{Jingkai Xia}\affiliation{\shKeyLab}
\author{Mengjiao Xiao}\affiliation{\UMD}
\author{Xiang Xiao}\affiliation{\SYU}
\author{Pengwei Xie}\affiliation{\TDLee}
\author{Binbin Yan}\affiliation{\TDLee}
\author{Xiyu Yan}\affiliation{\SYUzhuhai}
\author{Jijun Yang}\affiliation{\shKeyLab}
\author{Yong Yang}\affiliation{\shKeyLab}
\author{Yukun Yao}\affiliation{\shKeyLab}
\author{Chunxu Yu}\affiliation{\NKU}
%\author{Jumin Yuan}\affiliation{\SDUdep}\affiliation{\SDUlab}
\author{Ying Yuan}\affiliation{\shKeyLab}
\author{Zhe Yuan}\affiliation{\FDU} %
\author{Xinning Zeng}\affiliation{\shKeyLab}
\author{Dan Zhang}\affiliation{\UMD}
\author{Minzhen Zhang}\affiliation{\shKeyLab}
\author{Peng Zhang}\affiliation{\YaLongSD}
\author{Shibo Zhang}\affiliation{\shKeyLab}
\author{Shu Zhang}\affiliation{\SYU}
\author{Tao Zhang}\affiliation{\shKeyLab}
\author{Wei Zhang}\affiliation{\TDLee}
\author{Yang Zhang}\email[Corresponding author: ]{yangzhangsdu@email.sdu.edu.cn}\affiliation{\SDUdep}\affiliation{\SDUlab}
\author{Yingxin Zhang}\affiliation{\SDUdep}\affiliation{\SDUlab} %
\author{Yuanyuan Zhang}\affiliation{\TDLee}
\author{Li Zhao}\affiliation{\shKeyLab}
\author{Qibin Zheng}\affiliation{\USST}
\author{Jifang Zhou}\affiliation{\YaLongSD}
\author{Ning Zhou}\affiliation{\shKeyLab}\affiliation{\SJTUSC}
\author{Xiaopeng Zhou}\affiliation{\BUAA}
\author{Yong Zhou}\affiliation{\YaLongSD}
\author{Yubo Zhou}\affiliation{\shKeyLab}
\collaboration{PandaX Collaboration}
\noaffiliation

%% file: main.bbl
\begin{thebibliography}{10}

\bibitem{2009Diffuse}
Shunsaku Horiuchi, John~F. Beacom, and Eli Dwek.
\newblock Diffuse supernova neutrino background is detectable in
  {Super-Kamiokande}.
\newblock {\em Physical Review D}, 79(8):083013, 2009.

\bibitem{2020Stellar}
Daniel Kresse, Thomas Ertl, and Hans~Thomas Janka.
\newblock Stellar collapse diversity and the diffuse supernova neutrino
  background.
\newblock 2020.

\bibitem{PhysRevLett581494}
R.~M. Bionta, G.~Blewitt, C.~B. Bratton, D.~Casper, A.~Ciocio, R.~Claus,
  B.~Cortez, M.~Crouch, S.~T. Dye, S.~Errede, G.~W. Foster, W.~Gajewski, K.~S.
  Ganezer, M.~Goldhaber, T.~J. Haines, T.~W. Jones, D.~Kielczewska, W.~R.
  Kropp, J.~G. Learned, J.~M. LoSecco, J.~Matthews, R.~Miller, M.~S. Mudan,
  H.~S. Park, L.~R. Price, F.~Reines, J.~Schultz, S.~Seidel, E.~Shumard,
  D.~Sinclair, H.~W. Sobel, J.~L. Stone, L.~R. Sulak, R.~Svoboda, G.~Thornton,
  J.~C. van~der Velde, and C.~Wuest.
\newblock Observation of a neutrino burst in coincidence with supernova 1987a
  in the {Large Magellanic Cloud}.
\newblock {\em Phys. Rev. Lett.}, 58:1494--1496, Apr 1987.

\bibitem{PhysRevLett581490}
K.~Hirata, T.~Kajita, M.~Koshiba, M.~Nakahata, Y.~Oyama, N.~Sato, A.~Suzuki,
  M.~Takita, Y.~Totsuka, T.~Kifune, T.~Suda, K.~Takahashi, T.~Tanimori,
  K.~Miyano, M.~Yamada, E.~W. Beier, L.~R. Feldscher, S.~B. Kim, A.~K. Mann,
  F.~M. Newcomer, R.~Van, W.~Zhang, and B.~G. Cortez.
\newblock Observation of a neutrino burst from the supernova {SN1987A}.
\newblock {\em Phys. Rev. Lett.}, 58:1490--1493, Apr 1987.

\bibitem{ALEXEYEV1988209}
E.N. Alexeyev, L.N. Alexeyeva, I.V. Krivosheina, and V.I. Volchenko.
\newblock Detection of the neutrino signal from {SN 1987A} in the {LMC} using
  the {INR Baksan} underground scintillation telescope.
\newblock {\em Physics Letters B}, 205(2):209--214, 1988.

\bibitem{2016Real}
Real-time supernova neutrino burst monitor at {Super-Kamiokande}.
\newblock {\em Astroparticle Physics}, 81:39--48, 2016.

\bibitem{2017The}
M.G.~Aartsen et~al.
\newblock The {IceCube} realtime alert system.
\newblock {\em Astroparticle Physics}, 92:30--41, 2017.

\bibitem{2016Design}
Hanyu Wei, Logan Lebanowski, Fei Li, Zhe Wang, and Shaomin Chen.
\newblock Design, characterization, and sensitivity of the supernova trigger
  system at {Daya Bay}.
\newblock {\em Astroparticle Physics}, 2016.

\bibitem{2004SNEWS}
P.~Antonioli et~al.
\newblock S{NEWS: The SuperNova Early Warning System}.
\newblock {\em New Journal of Physics}, 6(1):114, 2004.

\bibitem{AlKharusi2021}
S~Al~Kharusi et~al.
\newblock S{NEWS} 2.0: a next-generation supernova early warning system for
  multi-messenger astronomy.
\newblock {\em New Journal of Physics}, 23(3):031201, 2021.

\bibitem{PhysRevD91389}
Daniel~Z. Freedman.
\newblock Coherent effects of a weak neutral current.
\newblock {\em Phys. Rev. D}, 9:1389--1392, Mar 1974.

\bibitem{article11}
Dmitry Akimov, J.~Albert, Peibo An, C.~Awe, Phil Barbeau, B.~Becker, Vladimir
  Belov, A.~Brown, A.~Bolozdynya, B.~Cabrera-Palmer, Mayra Cervantes,
  J.~Collar, Reynold Cooper, R.~Cooper, Clara Cuesta, D.~Dean, J.~Detwiler,
  A.~Eberhardt, Y.~Efremenko, and A.~Zderic.
\newblock Observation of {Coherent Elastic Neutrino-Nucleus Scattering}.
\newblock {\em Science}, 357:eaao0990, 08 2017.

\bibitem{PhysRevD105043008}
Anna~M. Suliga, John~F. Beacom, and Irene Tamborra.
\newblock Towards probing the diffuse supernova neutrino background in all
  flavors.
\newblock {\em Phys. Rev. D}, 105:043008, Feb 2022.

\bibitem{Aprile2020}
E.~Aprile et~al.
\newblock Projected {WIMP} sensitivity of the {XENONnT} dark matter experiment.
\newblock {\em Journal of Cosmology and Astroparticle Physics}, 2020(11):031,
  2020.

\bibitem{LZ2019sgr}
D.~S. Akerib et~al.
\newblock {The LUX-ZEPLIN (LZ) Experiment}.
\newblock {\em Nucl. Instrum. Meth. A}, 953:163047, 2020.

\bibitem{PhysRevLett127261802}
Yue~Meng et~al.
\newblock Dark {Matter Search Results from the PandaX-4T Commissioning Run}.
\newblock {\em Phys. Rev. Lett.}, 127:261802, Dec 2021.

\bibitem{Aalbers2016}
J.~Aalbers et~al.
\newblock D{ARWIN}: towards the ultimate dark matter detector.
\newblock {\em Journal of Cosmology and Astroparticle Physics}, 2016(11):017,
  2016.

\bibitem{PandaX:2024oxq}
Abdusalam Abdukerim et~al.
\newblock {PandaX-xT: a Multi-ten-tonne Liquid Xenon Observatory at the China
  Jinping Underground Laboratory}.
\newblock 2 2024.

\bibitem{PhysRevD85052007}
K.~Bays et~al.
\newblock Supernova relic neutrino search at {Super-Kamiokande}.
\newblock {\em Phys. Rev. D}, 85:052007, Mar 2012.

\bibitem{PhysRevC73035807}
John~F. Beacom and Louis~E. Strigari.
\newblock New test of supernova electron neutrino emission using {Sudbury
  Neutrino Observatory }sensitivity to the diffuse supernova neutrino
  background.
\newblock {\em Phys. Rev. C}, 73:035807, Mar 2006.

\bibitem{2016NCimR39M}
A.~{Mirizzi}, I.~{Tamborra}, H.~Th. {Janka}, N.~{Saviano}, K.~{Scholberg},
  R.~{Bollig}, L.~{H{\"u}depohl}, and S.~{Chakraborty}.
\newblock {Supernova neutrinos: production, oscillations and detection}.
\newblock {\em Nuovo Cimento Rivista Serie}, 39(1-2):1--112, February 2016.
\newblock The data can be made available upon request at
  https://wwwmpa.mpa-garching.mpg.de/ccsnarchive.

\bibitem{Hdepohl2014NeutrinosFT}
L.~H{\"u}depohl.
\newblock Neutrinos from the{ Formation, Cooling, and Black Hole Collapse of
  Neutron Stars}.
\newblock 2014.

\bibitem{Nakazato2013}
Ken'ichiro Nakazato, Kohsuke Sumiyoshi, Hideyuki Suzuki, Tomonori Totani,
  Hideyuki Umeda, and Shoichi Yamada.
\newblock S{UPERNOVA NEUTRINO LIGHT CURVES AND SPECTRA FOR VARIOUS PROGENITOR
  STARS: FROM CORE COLLAPSE TO PROTO-NEUTRON STAR COOLING}.
\newblock {\em The Astrophysical Journal Supplement Series}, 205(1):2, feb
  2013.
\newblock The data can be made available upon request at
  http://asphwww.ph.noda.tus.ac.jp/snn.

\bibitem{Dark}
HongGuang Zhang, Abdusalam Abdukerim, Wei Chen, Xun Chen, YunHua Chen, XiangYi
  Cui, BinBin Dong, DeQing Fang, ChangBo Fu, and Karl Giboni.
\newblock Dark matter direct search sensitivity of the {PandaX-4T} experiment.
\newblock {\em Science China(Physics,Mechanics $\&$ Astronomy)}, 62:031011,
  2019.

\bibitem{Yang_2022}
J.~Yang, X.~Chen, C.~He, D.~Huang, Y.~Huang, J.~Liu, X.~Ren, A.~Wang, M.~Wang,
  B.~Yan, K.~Yin, J.~Yang, Y.~Yang, and Q.~Zheng.
\newblock Readout electronics and data acquisition system of pandax-4t
  experiment.
\newblock {\em Journal of Instrumentation}, 17(02):T02004, feb 2022.

\bibitem{PhysRevD98030001}
M.~Tanabashi and K.~Hagiwara et~al.
\newblock Review {of Particle Physics}.
\newblock {\em Phys. Rev. D}, 98:030001, Aug 2018.

\bibitem{PhysRev1041466}
Richard~H. Helm.
\newblock Inelastic and {Elastic Scattering of 187-Mev Electrons from Selected
  Even-Even Nuclei}.
\newblock {\em Phys. Rev.}, 104:1466--1475, Dec 1956.

\bibitem{ENGEL1991114}
J.~Engel.
\newblock Nuclear form factors for the scattering of weakly interacting massive
  particles.
\newblock {\em Physics Letters B}, 264(1):114--119, 1991.

\bibitem{Kozynets2018SensitivitT}
Tetiana Kozynets, Scott Fallows, and Carsten~B. Krauss.
\newblock Sensitivity of the {PICO-500} bubble chamber to supernova neutrinos
  through coherent nuclear elastic scattering.
\newblock {\em Astroparticle Physics}, 2018.

\bibitem{Keil2003}
Mathias~Th. Keil, Georg~G. Raffelt, and Hans-Thomas Janka.
\newblock Monte{ Carlo Study of Supernova Neutrino Spectra Formation}.
\newblock {\em The Astrophysical Journal}, 590(2):971, jun 2003.

\bibitem{PhysRevD94}
Rafael~F. Lang, Christopher McCabe, Shayne Reichard, Marco Selvi, and Irene
  Tamborra.
\newblock Supernova neutrino physics with xenon dark matter detectors: A timely
  perspective.
\newblock {\em Phys. Rev. D}, 94:103009, Nov 2016.

\bibitem{PhysRevD97}
HuiLing Li, YuFeng Li, Meng Wang, LiangJian Wen, and Shun Zhou.
\newblock Towards a complete reconstruction of supernova neutrino spectra in
  future large liquid-scintillator detectors.
\newblock {\em Phys. Rev. D}, 97:063014, Mar 2018.

\bibitem{LATTIMER1991331}
James~M. Lattimer and F.~{Douglas Swesty}.
\newblock A generalized equation of state for hot, dense matter.
\newblock {\em Nuclear Physics A}, 535(2):331--376, 1991.

\bibitem{ABE201751}
K.~Abe et~al.
\newblock Detectability of galactic supernova neutrinos coherently scattered on
  xenon nuclei in {XMASS}.
\newblock {\em Astroparticle Physics}, 89:51--56, 2017.

\bibitem{WF_2023}
J.~Li et~al.
\newblock Waveform {Simulation in PandaX-4T}.
\newblock {\em preprint arXiv: 2312}.
\newblock 11072, 2013.

\bibitem{2018Noble}
M.Szydagis et~al.
\newblock Noble {Element Simulation Technique v2.0}.
\newblock 2018.

\bibitem{instruments5010013}
Matthew Szydagis, Grant~A. Block, Collin Farquhar, Alexander~J. Flesher,
  Ekaterina~S. Kozlova, Cecilia Levy, Emily~A. Mangus, Michael Mooney, Justin
  Mueller, Gregory R.~C. Rischbieter, and Andrew~K. Schwartz.
\newblock A {Review of Basic Energy Reconstruction Techniques in Liquid Xenon
  and Argon Detectors for Dark Matter and Neutrino Physics Using NEST}.
\newblock {\em Instruments}, 5(1), 2021.

\bibitem{PhysRevLett130021802}
Wenbo~Ma et~al.
\newblock Search for {Solar $^{8}\mathrm{B}$ Neutrinos in the PandaX-4T
  Experiment Using Neutrino-Nucleus Coherent Scattering}.
\newblock {\em Phys. Rev. Lett.}, 130:021802, Jan 2023.

\bibitem{Joyce2020}
Meridith Joyce, Shing-Chi Leung, László Molnár, Michael Ireland, Chiaki
  Kobayashi, and Ken’ichi Nomoto.
\newblock Standing on {the Shoulders of Giants: New Mass and Distance Estimates
  for Betelgeuse through Combined Evolutionary, Asteroseismic, and Hydrodynamic
  Simulations with MESA}.
\newblock {\em The Astrophysical Journal}, 902(1):63, oct 2020.

\bibitem{Feldman:1997qc}
Gary~J. Feldman and Robert~D. Cousins.
\newblock Unified approach to the classical statistical analysis of small
  signals.
\newblock {\em Phys. Rev. D}, 57:3873--3889, Apr 1998.

\end{thebibliography}
